\documentclass{ws-ijgmmp}
\begin{document}

\markboth{Steven Watterson}
{Chiral and flavour projection of Dirac-K\"{a}hler fermions}

\catchline{}{}{}{}{}
%
%%%%%%%%%%%%%%%%%%%%%%%%%%%%%%%%%%%%%%%%%%%%%%%%%%%%%%%%%%%%%%%%%%%%

\title{The chiral and flavour projection of Dirac-K\"{a}hler fermions in the geometric discretization
}

\author{STEVEN WATTERSON}

\address{Department of Mathematics, Trinity College, Dublin 2, Ireland and\\
Division of Pathway Medicine, University of Edinburgh Medical School, Chancellor's Building, 49 Little France Crescent, Edinburgh, EH16 4SB, Scotland.\\
\email{watterss@maths.tcd.ie}}

\address{}

\maketitle

\begin{history}
\received{(Day Month Year)}
\revised{(Day Month Year)}
\end{history}

\begin{abstract}
It is shown that an exact chiral symmetry can be described for Dirac-K\"{a}hler fermions using the two complexes of the geometric discretization.  This principle is extended to describe exact flavour projection and it is shown that this necessitates the introduction of a new operator and two new structures of complex.  To describe simultaneous chiral and flavour projection, eight complexes are needed in all and it is shown that projection leaves a single flavour of chiral field on each.
\end{abstract}


\section*{Introduction}
Differential geometry has proven to be highly valuable in extracting the geometric meaning of continuum vector theories.  Of particular interest has been the Dirac-K\"{a}hler formulation of fermionic field theory \cite{Kahler}, which uses the antisymmetry inherent in the product between differential forms to describe the Clifford algebra.  In order to regularize calculations, we are required to introduce a discrete differential geometry scheme and it would be ideal if this had the same properties as the continuum and the correct continuum limit.  However, defining such a scheme has proven to be very challenging.  

The difficulties are usually exhibited by the Hodge star, which maps a form to its complement in the space, and the wedge product between forms.  In a discretization, we would like the latter to allow the product rule to be satisfied and we would like both to be local.

Several discrete schemes have been proposed that address these difficulties with varying success.  Becher and Joos \cite{BJ, BJ2} used a lattice to define operators with many desirable properties, but that were non-local.  To resolve the non-locality, they introduced translation operators.  Kanamori and Kawamoto \cite{KK, KK2} also used a lattice and introduced a specific non-commutativity between the fields and discrete forms.  This allowed the product rule to be fulfilled, but they found that it became necessary to introduce a second orientation of form in order for their action to remain Hermitian.  Instead of a lattice, Balachandran {\it et al} \cite{bal1, ydri} used a quantized phase space to regularize their system, leading to a fuzzy coordinate space \cite{oconnor, martin}.  

In this paper, we shall build upon a proposal by Adams \cite{Adams} in
which he introduces two parallel lattices to maintain the locality of
the Hodge star and uses a finite element scheme to capture the
properties of the wedge product.  This proposal describes a local,
discrete differential geometry for an arbitrary topological space and
its formal aspects have been thoroughly studied by de Beauc\'{e}, Samik Sen, Siddartha Sen and Czech \cite{deBeauce, deBeauce3, deBeauce2, czech}.  However, here we want to focus on its application to the Dirac-K\"{a}hler formulation.

In lattice quantum chromodynamics (lattice QCD) calculations, it is common to see the staggered fermion
formulation used to describe fermions \cite{Kogut, KSB}.  This
formulation addresses the problem of fermion doubling \cite{Rothe} by
reducing the number of degenerate fermions to $2^{n/2}$ in $n$
dimensional space-time.  It is frequently used with the quarter-root
trick \cite{Davies, Aubin, Adams2} to provide a description
of a single fermion on the lattice, although this approach has
attracted some controversy \cite{Bunk, Peardon}.  The continuous
Dirac-K\"{a}hler formulation is regarded as providing the continuum
limit for the staggered fermion formulation and so a discrete
Dirac-K\"{a}hler formulation with continuum properties can potentially
offer great insight into how to develop non-degenerate, doubler-free
fermionic field theories for the lattice.  

In this paper, we show how the two lattices of Adams' proposal can be
used to describe chiral symmetry in the associated Dirac-K\"{a}hler
formulation.  We also see how the idea of using more than one lattice
can be extended to describe an exact flavour projection.  We find that
this necessitates the introduction of two new structures of lattice
and a new operator.  Finally, we evaluate the path integral for this
formulation, considering the effects of chiral and flavour projection.  This builds on our previous work \cite{me, me2, me3}.

\section*{Background: the Geometric Discretization}

\begin{figure}[ph]
\centerline{\psfig{file=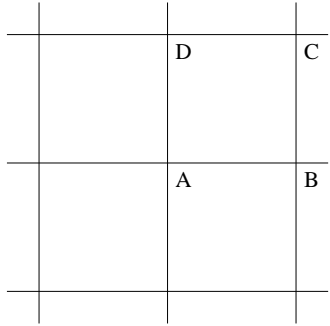,width=1.7in}}
\vspace*{8pt}
\caption{The complex in two dimensions.\label{twod}}
\end{figure}

Our starting point is the {\it complex}, which is the space on which we
define the discrete differential geometry.  It comprises the points of
the lattice, together with the links, faces, volumes and hyper-volumes
between the points.  Each point, link, face, volume or hyper-volume is
an example of a {\it simplex} and each simplex has an accompanying
cochain.  We denote a cochain by the
vertices of its corresponding simplex.  For example, we write the
cochain for the simplex between vertices $A$, $B$, $C$ and $D$ from
Fig. \ref{twod} as $[ABCD]$.  Each cochain is essentially a functional that acts upon a differential form of the same dimension as its simplex to give unity.  For example, $[ABCD]$ is defined such that 

\[
\int_{ABCD} dx^1 \wedge dx^2 = I \ .
\]

The cochains act as the discrete differential forms of the theory and a general field is a linear combination of cochains.  On the square $ABCD$, we write a general field as 
\[
\begin{array}{ccl}
\tilde{\Phi} & = & \tilde{\phi}([A])[A]+\tilde{\phi}([B])[B]+\tilde{\phi}([C])[C]+\tilde{\phi}([D])[D] \\
& & +\tilde{\phi}([AB])[AB]+\tilde{\phi}([DC])[DC]+\tilde{\phi}([DA])[DA] +\tilde{\phi}([CB])[CB] \\
& & +\tilde{\phi}([ABCD])[ABCD] \ .
\end{array} 
\]  

To define the wedge product between cochains, we must first introduce the Whitney map, which maps from the complex to the continuum, and the de Rham  map, which maps the other way.   

The Whitney map, $W$, replaces a cochain with a differential form of the same dimension as its accompanying simplex and introduces functions to interpolate in the regions between simplexes. For example, taking $ABCD$ to be a unit square with origin $A$, we introduce the interpolation functions
\[
\begin{array}{ccclcccl}
\mu_1(x) & = & (1-x_1) & \mbox{ for } x_1 \in ABCD \hspace{0.5cm} & \hspace{0.5cm} \mu_1(x) & = & 0 & \mbox{ otherwise} \\
\mu_2(x) & = & (1-x_2) & \mbox{ for } x_2 \in ABCD \hspace{0.5cm} & \hspace{0.5cm} \mu_2(x)  &  = & 0 & \mbox{ otherwise}
\end{array}
\]
where $x$ is the coordinate vector and this allows us to write 
\[
\begin{array}{l} 
W\left(\tilde{\phi}([A])[A]+\tilde{\phi}([B])[B]+\tilde{\phi}([C])[C]+\tilde{\phi}([D])[D]\right) = \tilde{\phi}([A])\mu_1(x)\mu_2(x) \\ 
\hspace{0.4cm} + \tilde{\phi}([B])(1-\mu_1(x))\mu_2(x) + \tilde{\phi}([C])(1-\mu_1(x))(1-\mu_2(x)) \\
\hspace{0.4cm} + \tilde{\phi}([D])\mu_1(x)(1-\mu_2(x)) \\
W\left(\tilde{\phi}([DA])[DA] + \tilde{\phi}([CB])[CB]+\tilde{\phi}([DC])[DC]+\tilde{\phi}([AB])[AB]\right) = \\
\hspace{0.4cm} \tilde{\phi}([DA])\mu_1(x) dx^2 + \tilde{\phi}([CB])(1-\mu_1(x))dx^2 + \tilde{\phi}([DC])(1-\mu_2(x))dx^1 \\
\hspace{0.4cm} + \tilde{\phi}([AB])\mu_2(x) dx^1\\
W\left(\tilde{\phi}([ABCD])[ABCD]\right) = \tilde{\phi}([ABCD])dx^1\wedge dx^2  .
\end{array} 
\]
The De Rham map, $R$, discretizes a field by integrating over the
simplexes whose dimension match that of the accompanying
differential form.  $R$ also introduces a cochain of the appropriate
dimension.  Thus,
\[
\begin{array}{cclccl}
\tilde{\phi}([A]) & = & \phi(x)|_{x=A} & \tilde{\phi}([B]) & = & \phi(x)|_{x=B}\\ 
\tilde{\phi}([C]) & = & \phi(x)|_{x=C} & \tilde{\phi}([D]) & = & \phi(x)|_{x=D}\\ 
\tilde{\phi}([DC]) & = & \int_{DC} \phi(x) dx^1 & \tilde{\phi}([AB]) & = & \int_{AB} \phi(x) dx^1\\
\tilde{\phi}([DA]) & = & \int_{DA} \phi(x) dx^2 & \tilde{\phi}([CB]) & = & \int_{CB} \phi(x) dx^2\\
\tilde{\phi}([ABCD]) & = & \int_{ABCD} \phi(x) dx^1\wedge dx^2 
\end{array}
\]
and
\[
\begin{array}{ccl}
R\left[\phi(x,\emptyset)\right] & = & \tilde{\phi}([A])[A] +
\tilde{\phi}([B])[B] + \tilde{\phi}([C])[C] + \tilde{\phi}([D])[D] \\
R\left[\phi(x,1)dx^1\right] & = &
\tilde{\phi}([DC])[DC]+\tilde{\phi}([AB])[AB] \\
R\left[\phi(x,2)dx^2\right] & = &
\tilde{\phi}([DA])[DA]+\tilde{\phi}([CB])[CB] \\
R\left[\phi(x,12)dx^1\wedge dx^2\right] & = &
\tilde{\phi}([ABCD])[ABCD] \ .
\end{array} 
\]

The wedge product between two discrete fields, $\tilde{\wedge}$, now takes the form
\begin{equation}
\label{wedgedef}
\tilde{\Phi} \tilde{\wedge} \tilde{\Theta} = R\left( W\left(\tilde{\Phi}\right) \wedge W\left(\tilde{\Theta} \right)\right)
\end{equation}
where $\wedge$ is the wedge product of the continuum.  We can take advantage of $W$ and $R$ to define the discrete exterior derivative as 
\[
D\tilde{\Phi} = R\left(d W\left(\tilde{\Phi}\right)\right) ,
\] 
where $d$ is the exterior derivative from the continuum, $dx^{\mu}\wedge \partial_{\mu}$.

In the continuum, the Hodge star is defined to be 
\[
*dx^H = \epsilon_{H, \mathcal{C}H} dx^{\mathcal{C}H} ,
\] 
where we have written $dx^H$ as shorthand for the product of forms $dx^{H_1}\wedge dx^{H_2}\wedge...\wedge dx^{H_h}$ and $\mathcal{C}H$ is the complement of $H$ in the space.  $\epsilon$ is the Levi-Civita tensor.  The square of the Hodge star has the property
\begin{equation}
\label{starsq}
**dx^H = (-1)^{h(n-h)}dx^H ,
\end{equation}
where $h$ is the dimension of the form and $n$ the dimension of the space.

\begin{figure}[h]
\centerline{\psfig{file=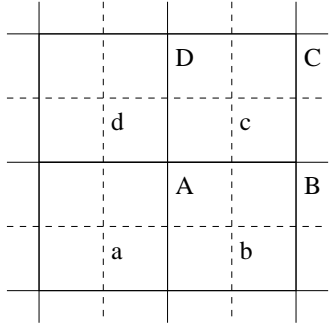,width=1.7in}}
\vspace*{8pt}
\caption{The original complex (solid lines) and its dual (dashed lines) in two dimensions.\label{origcomp}}
\end{figure}

To define the Hodge star discretely requires the introduction of a
second complex, known as the {\it dual}, in the same space as the
first. The dual (shown in Fig. \ref{origcomp} for two dimensions) is aligned with the original complex so that the mid-points of complementary simplexes coincide.  The Hodge star is defined so that it maps a cochain from one complex to a cochain from the other complex with an aligned simplex.  This gives the square of the operator, acting on a general cochain $[G]$, the following local form: $**[G] = (-1)^{g(n-g)}[G]$, where $g$ is dimension of the simplex and $n$ the dimension of the space.

With the Hodge star in place, the adjoint derivative can be defined as
\[
\delta [G] = (-1)^{ng+n+1}*D* [G]
\]
and the Laplacian can be written $(D-\delta)^2 = -D\delta - \delta D$, where $(D-\delta)$ is the Dirac-K\"{a}hler operator.

\begin{figure}[h]
\centerline{\psfig{file=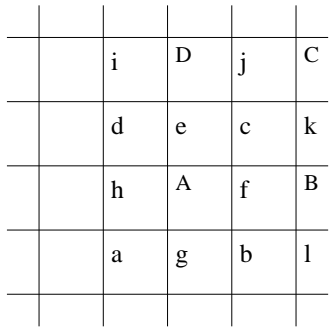,width=1.7in}}
\vspace*{8pt}
\caption{The barycentric subdivided complex in two dimensions.\label{bary}}
\end{figure}

To define the inner product, we must introduce the barycentric subdivided complex \cite{Senbary}.  The vertices of this complex
are formed from the midpoints of the simplexes on either the original or dual complex (the result is the same, whichever we choose).  These vertices are used in the construction of a new set of simplexes
and a new Whitney map, denoted $W^B$, which interprets a cochain
from either complex as a cochain on the
barycentric subdivided complex and maps it to a product of
differential forms and interpolating functions defined from the
simplexes of the barycentric subdivided complex.  The barycentric subdivided complex belonging to Fig. \ref{origcomp} is shown in Fig. \ref{bary}.  

This allows the inner product of two discrete fields, $\tilde{\Phi}$ and
$\tilde{\Omega}$, to be defined as
\[
<\tilde{\Phi}, \tilde{\Omega}> = \int W^B(\tilde{\Phi}) \wedge W^B(*\tilde{\Omega}) \ .
\]

\section*{Background: the Dirac-K\"{a}hler Basis}
In the Dirac-K\"{a}hler basis, the Clifford algebra is implemented
with the Clifford product, $\vee$, acting on differential forms,  
\[
\{dx^{\mu}, dx^{\nu}\}_{\vee} = 2\delta^{\mu\nu} \ .
\]
Formally, $\vee$ is defined to be
\[
\Phi(x) \vee \Theta(x) = \sum_{p\ge0}\frac{(-1)^{p \choose 2}}{p!}\left(\mathcal{A}^p e_{\mu_1}\lrcorner...e_{\mu_p}\lrcorner\Phi(x)\right) \wedge \left( e^{\mu_1}\lrcorner...e^{\mu_p}\lrcorner\Theta(x)\right) ,
\]
where $\mathcal{A}^p = (-1)^p$.  For a one-form, acting upon a general field 
\[
dx^{\mu} \vee \Phi(x) = \left(dx^{\mu} \wedge + e^{\mu} \lrcorner\right) \Phi(x) \ ,
\]
where $e^{\mu}\lrcorner$ is the contraction operator $e^{\mu}\lrcorner dx^H = \epsilon_{\mu, H/\mu}dx^{H/\mu}$ and $\Phi(x)$ is the linear combination of forms 
\[
\Phi(x) = \phi(x,\emptyset) + \sum_{\mu} \phi(x,\mu)dx^{\mu} + \sum_{\mu <\nu}\phi(x,\mu\nu)dx^{\mu}\wedge dx^{\nu}+... \ .
\]

The correspondence between the Dirac spinor, $\psi(x)$, and the Dirac-K\"{a}hler field, $\Phi(x)$, is established with  
\[
\Phi(x)=\sum_{ab} \psi(x)^{(b)}_a Z_{ab} \ ,
\]
where $Z$ is defined to be 
\[
Z = \sum_H (-1)^{h \choose 2} \gamma_H^{T} dx^H
\]
in Euclidean space-time.  $\gamma_H$ is shorthand for the product of the matrices $\gamma_{H_1}\gamma_{H_2}...\gamma_{H_h}$, where $\gamma_i$ take the form
\[
\gamma_j = i \left(\begin{array}{cc}
0 & \sigma_j \\
-\sigma_j & 0 \end{array}
\right) \hspace{1cm}
\gamma_4 = \left( \begin{array}{cc}
0 & 1 \\
1 & 0 \end{array}
\right)
\]
and $\sigma_j$ are the Pauli matrices.

The matrix $Z$ is pivotal to this correspondence because it has the properties
\[
dx^{\mu} \vee Z = \gamma_{\mu}^T Z \hspace{1cm} Z \vee dx^{\mu}  = Z \gamma_{\mu}^T ,
\]
which mean that
\begin{equation}
\label{corr1}
\begin{array}{ccl}
dx^{\mu}\vee\Phi(x) & = & \sum_{ab} \psi_a^{(b)}(x)\left(\gamma_{\mu}^TZ\right)_{ab} \\
 & = & \sum_{ab} \left(\gamma_{\mu} \psi(x)\right)_a^{(b)} Z_{ab}
\end{array}
\end{equation}
and
\begin{equation}
\label{corr2}
\begin{array}{ccl}
\Phi(x) \vee dx^{\mu} & = & \sum_{ab} \psi_a^{(b)}(x)\left(Z\gamma_{\mu}^T\right)_{ab} \\
& = & \sum_{ab} \left( \psi(x)\gamma_{\mu}\right)_a^{(b)} Z_{ab} \ .
\end{array}
\end{equation}

The components $\phi(x,H)$ and $\psi_a^{(b)}(x)$ are explicitly related by 
\begin{equation}
\label{equiv}
\begin{array}{cc}
\phi(x,H) = Tr\left(\gamma_H^{\dagger}\psi(x)\right) \qquad & \qquad \psi_a^{(b)} = \frac{1}{4}\sum_H \phi(x,H)\gamma^H_{ab} \ .
\end{array}
\end{equation}

On the complex, we find that the fields do not exhibit the properties
of Eqs. (\ref{corr1}) and (\ref{corr2}) exactly.  If this were
the case, we would find that, referring to Fig. \ref{twod}, 
\begin{equation}
\label{wedge}
\begin{array}{ccl}
\left([DC]+[AB]\right) & \tilde{\wedge} & \left(\tilde{\phi}([AD])[AD]+\tilde{\phi}([CB])[CB]\right) 
\\ & = & \int_{ABCD}dx^1 dx^2 Tr\left(\gamma_2\psi(x)\right)[ABCD] \ .
\end{array}
\end{equation}
However, in the right hand side of this expression, the integration is
over a domain of different dimension to that of the combination of
$\gamma$-matrices.  Using the definition of the de Rham map and
Eq. (\ref{equiv}), we can see that no such field exists in the
discretization, so, instead, the left hand side evaluates to 
\begin{equation}
\label{wedge2}
\begin{array}{ccl}
\left([DC]+[AB]\right) & \tilde{\wedge} & \left(\tilde{\phi}([AD])[AD]+\tilde{\phi}([CB])[CB]\right) \\
& = & \frac{1}{2}\left(\tilde{\phi}([DA])+\tilde{\phi}([CB])\right)[ABCD] \ ,
\end{array}
\end{equation}
which is a first order approximation to the right hand side of Eq. (\ref{wedge}).

In the continuum, the columns of the four by four matrix, $\psi(x)$, each correspond to a separate flavour of field which can be isolated using the flavour projection $\mathbf{P}^{(b)}\psi(x) = \psi(x)P^{(b)}$, where
\begin{equation}
\label{projector}
P^{(b)} = \frac{1}{4}\left(1 + i\alpha_b \gamma_1\gamma_2\right)\left(1+\beta_b\gamma_1\gamma_2\gamma_3\gamma_4\right)
\end{equation}
and
\[
\begin{array}{ccl}
\alpha_b & = & \left(-1, +1, -1, +1\right)^T \\ 
\beta_b & = & \left(-1, -1, +1, +1\right)^T .
\end{array}
\]

However, because the properties of Eqs. (\ref{corr1}) and (\ref{corr2}) are only approximately captured on the complex, we cannot use the discrete counterparts to 
\begin{equation}
\label{dxproj}
P^{(b)} = \frac{1}{4}\left(1+i\alpha_b dx^1\wedge dx^2\right) \vee \left(1+\beta_b dx^1\wedge dx^2 \wedge dx^3 \wedge dx^4\right)
\end{equation}
to facilitate flavour projection and $dx^1\wedge dx^2\wedge dx^3 \wedge dx^4$ to generate exact chiral symmetry.

However, we can take advantage of the relationship between the dual
and the original complexes to implement an exact chiral symmetry, as
we shall demonstrate in the next section.  In subsequent sections, we introduce additional complexes to implement exact flavour projection.

\section*{Chiral Symmetry}
It was shown by Rabin that, in the continuum, the chiral symmetry of Dirac-K\"{a}hler fields is related to the Hodge star \cite{Rabin}.  Using equation (\ref{equiv}), we can show that the substitution
\[
\psi(x)\rightarrow \gamma_5\psi(x)
\]
is equivalent to the transformation
\[
\Phi(x) \rightarrow -*\mathcal{BA}\Phi(x) \ ,
\]
where $\mathcal{B}$ and $\mathcal{A}$ are operators defined to be 
\[
\begin{array}{ccl}
\mathcal{B}dx^H & = & (-1)^{h \choose 2}dx^H \\
\mathcal{A}dx^H & = & (-1)^h dx^H .
\end{array}
\]

On the complex, the discrete fields are obtained from the continuous
fields.  As such, there is the
potential to use $*$ to describe chiral symmetry.
However, we cannot use the formulation as it stands, because the
fields associated with the simplexes from each complex are initially
discretized by integrating over different domains.  For example,
referring to Fig. \ref{origcomp}, $\tilde{\phi}([A])$ is obtained by
sampling $Tr\left(\psi(x)\right)$ at $A$ and $\tilde{\phi}([abcd])$ is
obtained by integrating
$Tr\left(\gamma_4\gamma_3\gamma_2\gamma_1\psi(x)\right)$ over $abcd$.
In this case, $\tilde{\Phi}\rightarrow-*\mathcal{BA}\tilde{\Phi}$ is not
equivalent to $\psi(x) \rightarrow \gamma_5\psi(x)$ because the
domains do not agree.  To attain this equivalence, we must modify the
domain of integration used to initially discretize the fields on one of
the complexes.  Whilst the choice is arbitrary, we will chose to modify the fields on the dual.

We introduce a new de Rham map, $R_0$, that is identical to $R$ on the
original complex, but that uses domains of integration on the dual that match simplices from the original complex.  It is defined so that fields on the
dual are discretized using domains of integration defined by the
simplexes from the original complex related to the simplexes of the
dual by their accompanying cochains and $*$.  Formally, $R_0$ is defined to be
\[
\begin{array}{ccll}
R_0[\phi(x,H)dx^H] & = & \sum_{H}\int_{H}\phi(x,H) dx^H & \hspace{0.5cm}\mbox{ on the original complex} \\
R_0[\phi(x,H)dx^H] & = & \sum_{H}\int_{\bar{*}H}\phi(x,H) dx^H & \hspace{0.5cm}\mbox{ on the dual complex} .
\end{array} 
\]
Here, $H$ is a simplex of the same space-time dimension as $dx^H$ and $\bar{*}H$ maps a simplex from the dual to its counterpart on the original complex obtained as the simplex associated with the cochain $\rho_{H,\mathcal{C}H}*[H]$.  We continue to
define the wedge product, Clifford product, exterior derivative and adjoint derivative using $R$.  Only for the initial discretization of the fields do we propose to use $R_0$.
 
With the fields discretized in this manner, we can generate an exact
chiral symmetry with the operator $-*\mathcal{BA}$ acting on
$\tilde{\Phi}$ which has the property 
\[
\{-*\mathcal{BA}, \left(D-\delta\right)\}=0 \ .
\]

To implement chiral projection, we deconstruct $*$ into $*_{od}$,
which is the Hodge star mapping cochains from the original complex to
the dual, and $*_{do}$ which is the Hodge star mapping cochains from the dual complex to the original.  This allows us to write chiral projection as
\[
P_{R/L} = \frac{1}{2}\left(1\mp*_{do}\mathcal{BA} \pm *_{od}\mathcal{BA}\right) .
\]
If we write $\hat{\tilde{\Phi}}=\tilde{\Phi}_o+\tilde{\Phi}_d$, where
$\tilde{\Phi}_o$ and $\tilde{\Phi}_d$ are the discrete fields on the
original and dual complexes, respectively, then
$P_R\hat{\tilde{\Phi}}$ will project the right handed degrees of
freedom of $\hat{\tilde{\Phi}}$ onto the original complex and the left
handed degrees of freedom onto the dual.

\section*{Flavour Symmetry}
The Dirac-K\"{a}hler field enjoys global $SU(4)$ flavour symmetry in
the continuum.  However, as can be seen from Eq. (\ref{dxproj}), flavour projection only requires the subgroups generated by $dx^1\wedge dx^2 \wedge dx^3 \wedge dx^4$ and $dx^1 \wedge dx^2$. 

As we have seen, on the complex, the n\"{a}ive implementation of these
symmetries is only approximate.  However, just as we were able to use
the dual and original complexes to describe an exact chiral symmetry,
we can use the dual and original complexes to describe 
the $dx^1\wedge dx^2\wedge dx^3\wedge dx^4$ symmetry needed for
flavour projection.

We write $\mathbf{P}^{(b)}$ as the product of two operators $\mathbf{P}^{(b)}_{\alpha}$ and $\mathbf{P}^{(b)}_{\beta}$ and $P^{(b)}$ as the product of the two projections
$P_{\beta}^{(b)}$ and $P_{\alpha}^{(b)}$, such that
\[
\begin{array}{ccl}
\mathbf{P}^{(b)}\Phi(x) & = & \mathbf{P}^{(b)}_{\beta}\mathbf{P}^{(b)}_{\alpha}\Phi(x) \\
& = & \Phi(x) \vee P^{(b)} \\
& = & \Phi(x) \vee P^{(b)}_{\alpha} \vee P^{(b)}_{\beta}
\end{array}
\]
where 
\[
\begin{array}{ccl}
P_{\beta}^{(b)} & = & \frac{1}{2}\left(1+\beta_b dx^1\wedge dx^2\wedge dx^3 \wedge dx^4\right) \\
P_{\alpha}^{(b)} & = & \frac{1}{2}\left(1+i\alpha_b dx^1 \wedge dx^2\right) ,
\end{array}
\]
In the continuum, we can show that
\begin{equation}
\label{flavourstar}
\Phi(x) \vee (dx^1 \wedge dx^2 \wedge dx^3 \wedge dx^4) = *\mathcal{B}\Phi(x) \ .
\end{equation}
and we can use this to write $\mathbf{P}^{(b)}_{\beta}$ as 
\begin{equation}
\label{projbeta}
\mathbf{P}^{(b)}_{\beta}\hat{\tilde{\Phi}} = \frac{1}{2}\left(1+\beta_b *_{do}\mathcal{B}-\beta_b*_{od}\mathcal{B}\right)\hat{\tilde{\Phi}} \ .
\end{equation}

Unfortunately, describing the symmetry generated by $dx^1 \wedge dx^2$
is more involved.  When we apply $dx^1\wedge dx^2 \wedge dx^3\wedge
dx^4$ to a general form, the form we obtain is complementary in all
four dimensions to the original and this allows us to describe this process in terms
of $*$.  If we consider applying $dx^1\wedge dx^2$ to a general form,
the form we obtain is complementary in the $\{1,2\}$ subspace, but equal to the original form in the $\{3,4\}$ subspace.    By analogy, we need an operator that maps a form to its complement in the $\{1,2\}$ subspace, but not in the $\{3,4\}$ subspace.

To this end we introduce $\spadesuit$, which, in the continuum, we
define to be 
\[
\spadesuit dx^H = \rho_{H_{12},\mathcal{C}_{12}H_{12}}dx^{\mathcal{C}_{12}H} .
\]   
Here, we have introduced $H_{12}$ as the components of $H$ belonging
to the $\{1,2\}$ subspace and $C_{12}$ as the complementary operator
in the $\{1,2\}$ subspace.  $\mathcal{C}_{12}H$ is equal to $H$ in the
$\{3,4\}$ subspace and complementary to $H$ in the $\{1,2\}$ subspace
and $C_{12}H_{12}$ is the complement of $H_{12}$ in the $\{1,2\}$ subspace.  
The square of $\spadesuit$ has the property

\[
\spadesuit\spadesuit dx^H = (-1)^{h_{12}(2-h_{12})}dx^H ,
\]
which is comparable to Eq. (\ref{starsq}).  With $\spadesuit$, we can show that
\[
\Phi(x) \vee(dx^1\wedge dx^2) = \spadesuit \mathcal{B}_{12}\Phi(x) \ ,
\]
where $\mathcal{B}_{12} dx^H = (-1)^{h_{12} \choose 2} dx^H$ and $h_{12}$ is the number of components of $H$ in the $\{1,2\}$ subspace.

To describe $\spadesuit$ in the discretization, we are required to
introduce a new complex, just as Adams was required to introduce the
dual to describe $*$.  The new complex should align with the original
complex so that simplexes that are complementary
in the $\{1,2\}$ subspace, but not the $\{3,4\}$ subspace, share midpoints.

\begin{figure}[h]
\centerline{\psfig{file=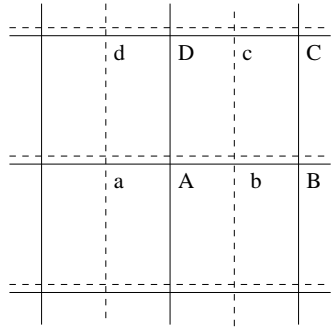,width=1.7in}}
\vspace*{8pt}
\caption{The original complex (solid lines) and the analogue of the 12c (dashed lines, shown slightly offset), in two dimensions.\label{12c}}
\end{figure}

We christen the new complex the $12c$ ($12$-complement) complex.  In
order to help visualize the alignment of the $12c$ complex, in Fig.
\ref{12c} we show its analogue in two dimensions.  Here, the simplexes with coincident midpoints are complementary to each other in the $1$ direction, but not the $2$ direction and we would define the analogue of $\spadesuit$ so that it maps between
the following pairs of cochains from $ABCD$ and $abcd$:-
\[
\begin{array}{ccc}
[A] & \leftrightarrow & [ab] \\
{[D]} & \leftrightarrow & [dc] \\
{[c]} & \leftrightarrow & [DC] \\
{[b]} & \leftrightarrow & [AB] \\
{[DA]} & \leftrightarrow & [abcd] \\
{[cb]} & \leftrightarrow & [ABCD] \ .
\end{array}
\]      

It is worth mentioning that, in a two dimensional theory, we would not need to introduce this complex to isolate the flavours because the origianl and dual complexes would be sufficient.  It is simply an analogue to the $12c$ complex.

\begin{figure}[h]
\centerline{\psfig{file=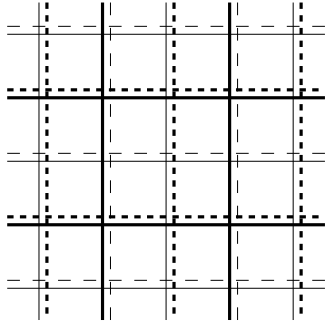,width=1.7in}}
\vspace*{8pt}
\caption{The original complex (light, solid lines), the dual (bold, solid lines) and the analogues of the 12c (light, dashed lines, offset) and the 12cd (bold, dashed lines, offset), in two dimensions.\label{12cd}}
\end{figure}

In order to complete the description of flavour projection,
we must go further and define a fourth complex.  The term proportional
to $(dx^1 \wedge dx^2) \vee (dx^1\wedge dx^2 \wedge dx^3 \wedge dx^4)$
in Eq. (\ref{dxproj}) maps a form to its complement in the
$\{1,2\}$ subspace, before mapping the resulting form to its
complement in all four dimensions.  The end result is a form complementary to the original in the $\{3,4\}$ subspace, but not in the $\{1,2\}$ subspace.  In the continuum, this can be described as a combination of $\spadesuit$ and $*$, but to capture
this map discretely requires us to introduce a fourth complex.  The
fourth complex is the dual of the $12c$ complex (or the
$12$-complement of the dual, it can be viewed either way) and we
christen it the $12cd$ ($12$ complement's dual) complex.  The two
dimensional analogue of the $12cd$ complex, together with the original, dual and the analogue of the $12c$ complex, is shown in Fig. \ref{12cd}.

To enable flavour projection between these four complexes, we must
ensure that the fields are initially discretized using compatible domains of integration.  To this end, we extend the definition of $R_0$ so that the fields on all four complexes are initially discretized using domains of integration taken from the original complex. $R_0$ becomes
\[
\begin{array}{ccll}
R_0[\phi(x,H)dx^H] & = & \sum_H \int_{H} \phi(x,H) dx^H & \hspace{0.5cm}\mbox{ on the original complex} \\
R_0[\phi(x,H)dx^H] & = & \sum_H \int_{\bar{*}H} \phi(x,H) dx^H & \hspace{0.5cm}\mbox{ on the dual complex} \\
R_0[\phi(x,H)dx^H] & = & \sum_H \int_{\bar{\spadesuit}H} \phi(x,H)dx^H & \hspace{0.5cm}\mbox{ on the $12c$ complex and} \\
R_0[\phi(x,H)dx^H] & = & \sum_H \int_{\bar{*}\bar{\spadesuit}H} \phi(x,H)dx^H & \hspace{0.5cm}\mbox{ on the $12cd$ complex} ,
\end{array}
\]
where $\bar{*}$ is defined as before and $\bar{\spadesuit}$ maps a simplex to its complement in the $\{1,2\}$ subspace, but not the $\{3,4\}$ subspace.  Formally, $\bar{\spadesuit}H$ generates the simplex associated with the cochain generated by $\rho_{H_{12},\mathcal{C}_{12}H_{12}}\spadesuit [H]$.

We must also extend the definition of $\mathbf{P}^{(b)}_{\beta}$ from Eq.
(\ref{projbeta}) to include the maps between the $12c$ and $12cd$ complexes.  If we rewrite $\hat{\tilde{\Phi}}$ as
\[
\hat{\tilde{\Phi}} = \tilde{\Phi}_o + \tilde{\Phi}_d + \tilde{\Phi}_e + \tilde{\Phi}_t \ ,
\]
where $\tilde{\Phi}_e$ and $\tilde{\Phi}_t$ are the contributions from the $12c$ and $12cd$ complexes, respectively, $\mathbf{P}_{\beta}^{(b)}$ now takes the form
\[
\begin{array}{ccl}
\mathbf{P}^{(b)}_{\beta} & = & \frac{1}{2}\left(1+\beta_b*_{do}\mathcal{B} - \beta_b*_{od}\mathcal{B} \right. \\
& & \left. + \beta_b*_{te}\mathcal{B} - \beta_b*_{et}\mathcal{B}\right) ,
\end{array}
\]
where $*_{te}$ maps from the $12cd$ complex to the $12c$ complex and $*_{et}$ maps the other way.

To describe $\mathbf{P}^{(b)}_{\alpha}$, we similarly deconstruct $\spadesuit$
into $\spadesuit_{eo}$ which maps from the $12c$ complex to the
original, $\spadesuit_{oe}$, which maps the other way,
$\spadesuit_{td}$, which maps from the $12cd$ complex to the dual and $\spadesuit_{dt}$ which maps the other way.  This allows us to write $\mathbf{P}^{(b)}_{\alpha}$ as
\[
\mathbf{P}^{(b)}_{\alpha} = \frac{1}{2}\left(1+i \alpha_b \spadesuit_{eo}\mathcal{B}_{12} -i\alpha_b \spadesuit_{oe}\mathcal{B}_{12} +i\alpha_b \spadesuit_{td}\mathcal{B}_{12}-i\alpha_b \spadesuit_{dt}\mathcal{B}_{12}\right)
\]
and flavour projection can now be written as $\mathbf{P}^{(b)}\hat{\tilde{\Phi}} = \mathbf{P}^{(b)}_{\beta}\mathbf{P}^{(b)}_{\alpha}\hat{\tilde{\Phi}}$.

One of the properties of the geometric discretization is that the Dirac-K\"{a}hler operator maps the degrees of freedom from each complex in the same way.  For example, referring to $ABCD$ and $abcd$ from Fig. \ref{origcomp}, $(D-\delta)$ maps $\tilde{\phi}([ABCD])[ABCD]$ to $\frac{1}{2}\tilde{\phi}([ABCD])[DA]$ and this is matched by the behaviour of $(D-\delta)$ on the dual which maps $\tilde{\phi}([c])[c]$ to $\frac{1}{2}\tilde{\phi}([c])[dc]$. Consequently, the cancellation properties of $\tilde{\phi}([ABCD])$ and $\tilde{\phi}([c])$ will be equally valid before and after the application of $(D-\delta)$, so we have  
\[
[\mathbf{P}^{(b)}, (D-\delta)]\hat{\tilde{\Phi}}=0 \ .
\]

To illustrate $\mathbf{P}^{(b)}$, we shall
consider the effect of $\mathbf{P}^{(1)}$ on $\hat{\tilde{\Phi}}$.  In
particular, we will consider $\mathbf{P}^{(1)}$ in stages, so first we apply
$\mathbf{P}^{(1)}_{\alpha}$ to $\hat{\tilde{\Phi}}$.  This leaves the degrees of freedom belonging to the first and third columns of $\psi$ on the ordinary and dual complexes and the degrees of freedom belonging to the second and fourth columns of $\psi$ on the $12c$ and $12cd$ complexes.  If we now apply $\mathbf{P}^{(1)}_{\beta}$ to this system, it will project between the original and dual complexes to leave the degrees of freedom belonging to the first column of $\psi$ on the original complex and the degrees of freedom belonging to the third column of $\psi$ on the dual complex.  Between the $12c$ and $12cd$ complexes, $\mathbf{P}^{(1)}_{\beta}$ will leave the degrees of freedom belonging to the second column of $\psi$ on the $12c$ complex and the degrees of freedom belonging to the fourth column of $\psi$ on the $12cd$ complex.

\section*{\label{sim}Simultaneous Chiral and Flavour Projection}
In the previous two sections, we showed that it was possible to
implement exact chiral symmetry using the original and dual complexes
and flavour projection using the original, dual, $12c$ and $12cd$
complexes.  However, because both projections use the original and the
dual complexes in different ways, we cannot implement chiral and flavour projection simultaneously using the
formulation as it stands.  

To illustrate this point, we consider
$\mathbf{P}^{(1)}P_{R}\hat{\tilde{\Phi}}$, which we write as
$\mathbf{P}^{(1)}_{\beta}\mathbf{P}^{(1)}_{\alpha}P_R\hat{\tilde{\Phi}}$.  For the
four complexes, $P_{R/L}$ becomes
\[
P_{R/L} = \frac{1}{2}\left(1\mp *_{do}\mathcal{BA}\pm
*_{od}\mathcal{BA}\mp *_{te}\mathcal{BA} \pm *_{et}\mathcal{BA} \right) .
\]

$P_R\hat{\tilde{\Phi}}$ leaves the degrees of freedom belonging to the
upper components of $\psi$ on the original and $12c$ complexes and the
degrees of freedom belonging to the lower components of $\psi$ on the
dual and $12cd$ complexes.  To this, we apply $\mathbf{P}^{(1)}_{\alpha}$,
which projects between the original and $12c$ complexes to leave the
degrees of freedom belonging to the upper components of the first and third columns of $\psi$ on the original complex and the degrees of freedom
belonging to the upper components of the second and fourth columns of $\psi$ on
the $12c$ complex.  $\mathbf{P}^{(1)}_{\alpha}$ also projects between the dual
and $12cd$ complexes to leave the degrees of freedom belonging to the
lower components of the first and third columns on the dual complex and the
degrees of freedom belonging to the lower components of the second and fourth columns of $\psi$ on the $12cd$ complex.  If we now consider applying
$\mathbf{P}^{(1)}_{\beta}$ to this system, we see that $\mathbf{P}^{(1)}_{\beta}$
projects between the original and dual complexes to leave the degrees
of freedom belonging to both the upper and lower components of the first and third columns
on both complexes.  It also projects between the $12c$ and
$12cd$ complexes to leave the degrees of freedom belonging to both the upper and lower components of the second and fourth columns on both complexes.  

Because $\mathbf{P}^{(b)}_{\beta}$ and $P_{R/L}$ both map between the original and dual complexes and the $12c$ and $12cd$ complexes in different ways, we cannot use these definitions for $\mathbf{P}^{(b)}$ and $P_{R/L}$ to isolate non-degenerate, chiral Dirac-K\"{a}hler fermions.

However, we can overcome this difficulty, by introducing a second set
of complexes that are duplicates of the existing four.  By
introducing a second set, we can redefine the chiral projection, so that it maps between complexes
from different sets, whilst continuing to define the flavour projection so that it maps
between complexes from the same set.  
This arrangement allows us to use $P_{R/L}$ to place only the degrees
of freedom belonging to the right handed components of $\psi$ on one set of complexes and the degrees of freedom belonging to the left
handed components of $\psi$ on the other.  The flavour projection
within each set will now no longer mix different chiralities of field.

To formally define this system, we label
the first set of complexes $A$ and the duplicate set $B$.  We write the Hodge star operator in
the form $*^{XY}_{ab}$, where $XY$ labels the sets from which and to
which the Hodge star
maps and $ab$ label the complexes from which and to which it
maps, respectively.  Because $\spadesuit$ only maps between complexes of one set, it is unnecessary to modify its notation.  

The chiral projection operator now takes the form
\[
\begin{array}{ccl}
P_{R/L} & = & \frac{1}{2}\left( 1\mp *^{BA}_{do}\mathcal{BA}\mp
*^{BA}_{od}\mathcal{BA}\mp*^{BA}_{et}\mathcal{BA}\mp *^{BA}_{te}\mathcal{BA}\right.
\\
& & \left.\pm *^{AB}_{do}\mathcal{BA} \pm *^{AB}_{od}\mathcal{BA}\pm
*^{AB}_{et}\mathcal{BA} \pm *^{AB}_{te}\right)
\end{array}
\]
and we write $\hat{\tilde{\Phi}}=\tilde{\Phi}_o^A+\tilde{\Phi}_d^A+\tilde{\Phi}_e^A+\tilde{\Phi}_t^A+\tilde{\Phi}_o^B+\tilde{\Phi}_d^B+\tilde{\Phi}_e^B+\tilde{\Phi}_t^B$.  $P_R \hat{\tilde{\Phi}}$ places the degrees of freedom belonging to the right handed components of $\psi$ on the complexes of set $A$ and the degrees of freedom belonging to the left handed components of $\psi$ on the complexes of set $B$.

The flavour projection operator is now $\mathbf{P}^{(b)}=\mathbf{P}^{(b)}_{\beta}\mathbf{P}^{(b)}_{\alpha}$, where
\[
\begin{array}{ccl}
\mathbf{P}^{(b)}_{\beta} & = & \frac{1}{2}\left(1+\beta_b*^{AA}_{do}\mathcal{B}
- \beta_b*^{AA}_{od}\mathcal{B} + \beta_b*^{AA}_{te}\mathcal{B} -
\beta_b*^{AA}_{et}\mathcal{B}\right. \\
& & \hspace{0.4cm} \left.+\beta_b*^{BB}_{do}\mathcal{B}
- \beta_b*^{BB}_{od}\mathcal{B} + \beta_b*^{BB}_{te}\mathcal{B} - \beta_b*^{BB}_{et}\mathcal{B}\right) \\
\mathbf{P}^{(b)}_{\alpha} & = & \frac{1}{2}\left(1+i \alpha_b \spadesuit_{eo}\mathcal{B}_{12} -i\alpha_b
\spadesuit_{oe}\mathcal{B}_{12} +i\alpha_b \spadesuit_{td}\mathcal{B}_{12}  -i\alpha_b
\spadesuit_{dt}\mathcal{B}_{12}\right) .  \\
\end{array}
\]

As before, we describe the application of
$\mathbf{P}^{(1)}=\mathbf{P}^{(1)}_{\beta}\mathbf{P}^{(1)}_{\alpha}$ to $P_R\hat{\tilde{\Phi}}$
in stages.  $\mathbf{P}^{(1)}_{\alpha}P_R\hat{\tilde{\Phi}}$ leaves the degrees
of freedom belonging to the upper components of the first and third columns of
$\psi$ on the original and dual complexes of set $A$ and the lower components of the first and third columns on
the original and dual complexes of set $B$.  It also leaves the
degrees of freedom belonging to the upper components of the second and fourth columns of $\psi$ on the $12c$ and $12cd$ complexes of set $A$ and the
lower components of the second and fourth columns of $\psi$ on the $12c$ and
$12cd$ complexes of set $B$.

Applying $\mathbf{P}^{(1)}_{\beta}$ to this system will leave the degrees of
freedom belonging to the upper components of the first column of $\psi$ on
the original complex of set $A$ and the lower components of the first column of $\psi$ on the original complex of set $B$.  It will also leave the
degrees of freedom belonging to the upper components of the third column of
$\psi$ on the dual complex of set $A$ and the lower components of the third
column of $\psi$ on the dual complex of set $B$.  Similarly, it
will leave the degrees of freedom belonging to the upper components of the second
column of $\psi$ on the $12c$ complex of set $A$ and the lower
components of the second column of $\psi$ on the $12c$ complex of set $B$.
Lastly, it will also leave the degrees of freedom belonging to the
upper components of the fourth column of $\psi$ on the $12cd$ complex of set
$A$ and the lower components of the fourth column of $\psi$ on the $12cd$
complex of set $B$.

Using these definitions, $\mathbf{P}^{(b)}$ and $P_{R/L}$ can now be used to isolate a non-degenerate, chiral Dirac-K\"{a}hler fermion on each complex.

\section*{Path Integral Formulation}
The flavour and chiral projections have important consequences for
the path integral.  If we initially consider only flavour projection, the action must include contributions from all four complexes.  
\[
S = S_o+S_d+S_{12c}+S_{12cd} \ ,
\]
where, on each complex, we have
\[
S_i = <\tilde{\bar{\Phi}}_i, (D-\delta)\tilde{\Phi}_i> \ .
\]    
$\psi$ contributes different degrees of freedom to the fields on each complex whose cochains are related by $\spadesuit$ and $*$, so we can write the path integral as the product of four path integrals
\[
Z=\prod_{i=\mbox{orig},\mbox{dual},\mbox{12c},\mbox{12cd}}\left(\int [d\tilde{\bar{\Phi}}_i][d\tilde{\Phi}_i]e^{-S_i(\tilde{\bar{\Phi}}_i,\tilde{\Phi}_i)}\right)^{L_0} ,
\]
where $L_0$ is an as yet undefined constant.

Because each complex lies in a separate space, we can evaluate each
contribution separately and, on each complex, we have the
standard result
\[
\int [d\tilde{\bar{\Phi}}_i][d\tilde{\Phi}_i]\left(e^{-<\tilde{\bar{\Phi}}_i,
  (D-\delta)\tilde{\Phi}_i>}\right)^{L_0} = det\left[(D-\delta)\right]^{L_0} ,
\]
where $(D-\delta)$ is defined as a matrix operator.  Consequently, the full path integral takes the form
\[
Z = \left[det\left(D-\delta\right)\right]^{4L_0}
\]
and for this to have the correct continuum limit, we require that $L_0 = \frac{1}{4}$.  Interestingly, this means that the form of the path integral on each complex, prior to flavour projection, is
\[
\int [d\tilde{\bar{\Phi}}_i][d\tilde{\Phi}_i]\left(e^{-<\tilde{\bar{\Phi}}_i,
  (D-\delta)\tilde{\Phi}_i>}\right)^{\frac{1}{4}} ,
\]
which is exactly equivalent to that used in the quarter-root trick of lattice QCD \cite{Davies, Aubin, Adams2, Bunk, Peardon}.  

Applying flavour projection to this system changes the relationship between $\tilde{\Phi}$ and $\psi$ so that $\tilde{\bar{\Phi}}_i$ and $\tilde{\Phi}_i$ each describe the degrees of freedom belonging to one column of $\bar{\psi}$ and $\psi$, respectively.  This applies to the measures of integration as well as the fields in the action.  One consequence of this projection is that several fields on each complex come to share the same structure in terms of $Tr$, $\gamma_i$ and $\psi$.  For example, after flavour projection, referring to Fig. \ref{twod}, 
\[
\tilde{\phi}_o([A])=\psi(x)_1^{(1)}|_{x=A}, \hspace{1cm} \tilde{\phi}_o([ABCD])=-\int_{ABCD}dx^1\wedge dx^2 \psi(x)_1^{(1)} \ .
\]
However, because these fields are integrated over different domains, they are independent and the path integral is evaluated in the same way to give $Z=det(D-\delta)$.

If we now extend the path integral so that the action permits chiral projection, we must include the contribution from both sets of complexes.  The action becomes
\[
S = S^A_o+S^A_d+S^A_{12c}+S^A_{12cd}+S^B_o+S^B_d+S^B_{12c}+S^B_{12cd}
\]
and, prior to flavour projection, we can evaluate the path integral separately on each complex to obtain
\[
Z = [det\left(D-\delta\right)]^{8L_0} ,
\]
which requires us to set $L_0=\frac{1}{8}$, in order to obtain the correct continuum limit.  Applying both chiral and flavour projection to this system, as in the previous section, changes the relationship between $\tilde{\Phi}$ and $\psi$ so that the path integral comes to be a product of four, single flavour, right-handed path integrals and four, single flavour, left-handed path integrals.  As in the previous case, fields sharing a similar structure of $Tr$, $\gamma_i$ and $\psi$ are independent because their domains of integration differ.  However, because the components of $\hat{\tilde{\Phi}}$ on each complex can be unambiguously described as left or right handed, the chiral projection sends half the components to zero.  Consequently, the effective dimension of $(D-\delta)$ is halved on each complex.

\section*{Conclusion}
Here we have shown that it is possible to describe exact chiral symmetry for Dirac-K\"{a}hler fermions using the two complexes of the geometric discretization.  We have extended this idea to describe exact flavour projection and we have shown that this necessitated the introduction of two new structures of complex as well as a new operator.  To describe simultaneous chiral and flavour projection, we introduced a duplicate set of complexes and we were required to carefully define chiral projection so that it operates between sets and flavour projection so that it operates within sets.  This allowed us to project a single flavour of chiral field onto each complex.

We have observed that evaluating the path integral on each of the four complexes, prior to flavour projection and without the provision for simultaneous chiral projection, leads to a form equivalent to that used in the quarter-root trick of lattice QCD.

\end{document}